\documentclass[epj]{svjour}
\newcommand\rf[1]{(\ref{eq:#1})}
\newcommand\lab[1]{\label{eq:#1}}
\newcommand\nonu{\nonumber}
\newcommand\br{\begin{eqnarray}}
\newcommand\er{\end{eqnarray}}
\newcommand\be{\begin{equation}}
\newcommand\ee{\end{equation}}

\newcommand\lcurl{\left\{}
\newcommand\rcurl{\right\}}
\renewcommand\({\left(}
\renewcommand\){\right)}
\newcommand\bc{\begin{center}}
\newcommand\ec{\end{center}}

\newcommand\partder[2]{{{\partial {#1}}\over{\partial {#2}}}}
\newcommand\Sbr[2]{\Bigl\lbrack\,{#1}\, ,\,{#2}\,\Bigr\rbrack} 
\renewcommand\a{\alpha}
\renewcommand\b{\beta}

\renewcommand\l{\lambda}
\renewcommand\L{\Lambda}

\renewcommand\o{\over}

\newcommand\p{\phi}
\newcommand\vp{\varphi}
\renewcommand\P{\Phi}
\newcommand\pa{\partial}

\renewcommand\t{\tau}
\renewcommand\th{\theta}

\newcommand\twomat[4]{\left(\begin{array}{cc}  %%   2x2 matrix  %ESA
{#1} & {#2} \\ {#3} & {#4} \end{array} \right)}

\newcommand\cA{{\mathcal A}}
\newcommand\cB{{\mathcal B}}

\newcommand\cD{{\mathcal D}}

\newcommand\cF{{\mathcal F}}
\newcommand\cG{{\mathcal G}}

\newcommand\cL{{\mathcal L}}

\newcommand\cW{{\mathcal W}}

\newcommand\one{\hbox{{1}\kern-.25em\hbox{l}}}
\newcommand\0[1]{\relax\ifmmode\mathaccent"7017{#1}%
        \else\accent23#1\relax\fi}

\newcommand\DB{{Darboux-B\"{a}cklund}~}

\newcommand\st[2]{\stackrel{(#1 )}{#2}}

\newcommand\cSKP{$\textsl{SKP}^{N=1}_{(M_B,M_F)}$~}
\newcommand\cSKPe{$\textsl{SKP}^{N=2}_{(M_B,M_F)}$~}

\newcommand{\ct}[1]{\cite{#1}}
\newcommand{\bi}[1]{\bibitem{#1}}
\newcommand\NPB[3]{\textsl{Nucl. Phys.} \textbf{B#1}, #3 (#2)}
\newcommand\CMP[3]{\textsl{Commun. Math. Phys.} \textbf{#1}, #3 (#2)}

\newcommand\JMP[3]{\textsl{J. Math. Phys.} \textbf{#1}, #3 (#2)}

\newcommand\IJMPA[3]{\textsl{Int. J. Mod. Phys.} \textbf{A#1}, #3 (#2)}

\begin{document}
\title{Properties of Supersymmetric Integrable Systems of KP Type}
\author{E. Nissimov \and S. Pacheva% etc
% \author{E. Nissimov\inst{1} \and S. Pacheva\inst{2}% etc
}                     % Do not remove
\institute{Institute for Nuclear Research and Nuclear Energy, Boul.
Tsarigradsko Chausse 72, BG-1784 Sofia, Bulgaria\\
nissimov@inrne.bas.bg , svetlana@inrne.bas.bg}
\date{Received: date / Revised version: date}
% The correct dates will be entered by Springer
%
\abstract{The recently proposed supersymmetric extensions of reduced
Kadomtsev-Petviashvili (KP) integrable hierarchies in $N\!\! =\!\! 1,2$ 
superspace are shown to contain in the purely bosonic limit new types of 
ordinary non-supersymmetric integrable systems. The latter are coupled systems
of several multi-component non-linear Schr{\"o}dinger-like hierarchies whose 
basic nonlinear evolution equations contain additional quintic and
higher-derivative nonlinear terms. Also, we obtain the $N\!\! =\!\! 2$
supersymmetric extension of Toda chain model as Darboux-B{\"a}cklund
orbit of the simplest reduced $N\!\! =\!\! 2$ super-KP hierarchy and find its
explicit solution.
% Also the general Darboux-B{\"a}cklund (``supersymmetric soliton''-like)
% solutions for the pertinent super-KP hierarchies are explicitly given.
%
 % end of PACS codes
} %end of abstract
\maketitle
%
% Don't forget to give each section
% and subsection a unique label (see Sect.~\ref{sec:1}).
%
\section{Introduction}
\label{intro}
The notion of {\em integrable systems} arises in different disguises in a 
vast array of actively developing topics of theoretical physics. The last decade
has witnessed a dramatic increase in the interest towards integrable 
hierarchies of nonlinear evolution (``soliton'' or ``soliton-like'') equations,
especially towards their {\em supersymmetric} extensions, which is primarily 
due to the role they are playing in modern superstring theory. In theoretical 
physics {\em supersymmetry} is a fundamental symmetry principle unifying
bosonic and fermionic degrees of freedom of infinite-dimensional dynamical 
(field-theoretic) systems which underly superstring theory as an 
ultimate candidate for an unified theory of all fundamental forces in Nature,
including quantum gravity. In particular, supersymmetric generalizations of 
Kadomtsev-Petviashvili (super-KP) integrable hierarchy have been found to be of
direct relevance for {\em random matrix} models of non-perturbative 
superstring theory \ct{SI-sstring}. Supersymmetric integrable systems attract 
a lot of interest also from purely mathematical point of view, in particular,
the supersymmetric generalizations of the inverse scattering method, 
bi-Hamiltonian structures, tau-functions and Sato Grassmannian approach, and 
the Drinfeld-Sokolov agebraic scheme.

The purpose of the present contribution is to study in some detail the
properties of the recently proposed classes \cSKP and \cSKPe of reduced super-KP
integrable hierarchies \ct{match,N2-SKP-other,svirflow,n2-skp}
(see Eqs.\rf{Lax-SKP-N2},\rf{SKP-N2-isospec-eqs} and
Eqs.\rf{Lax-SKP-N1}--\rf{SKP-N1-isospec-eqs} below). We will show that the
latter contain in the purely bosonic limit new types of ordinary
non-supersymmetric integrable systems. Furthermore, we will show that the
supersymmetric extension of Toda chain in $N\!\! =\!\! 2$ superspace naturally 
arises as \DB orbit of the simplest member of \cSKPe class similarly to the
simpler $N\!\! =\!\! 1$ super-KP case \ct{match}. Thus, one can expect that the 
super-tau functions of the simplest members of \cSKP and \cSKPe integrable 
hierarchies will play, under certain additional constraints on them, the role 
of partition functions in random matrix models of superstrings similarly to the
case of random matrix models of ordinary non-supersymmetric strings (for a
review, see \ct{Morozov-UFN}).

\section{Sato Formulation of Super-KP Hierarchies}
\label{sec:1}
We shall use throughout the supersymmetric extension of Sato pseudo-differential
operator formalism in $N\!\! =\!\! 2$ superspace \ct{N2-SKP-other,n2-skp} with 
coordinates $(x,\th_{+},\th_{-})$, where $\th_{\pm}$ are anticommuting, and 
with the following standard notations for the ordinary (bosonic) derivative 
$\pa \equiv \partder{}{x}$ and the two super-covariant fermionic derivative 
$\cD_{\pm}$ operators: 
\be
\cD_{\pm} = \partder{}{\th_{\pm}} + \th_{\pm} \partder{}{x} \quad, \quad
\cD_{\pm}^2 = \pa \quad ,\quad \lcurl \cD_{+},\,\cD_{-}\rcurl = 0
\lab{N2-notat}
\ee
Any $N\!\! =\!\! 2$ super-pseudo-differential operator $\cA$ has the general form
$\cA = \cA_{+} + \cA_{-}$ with:
\be
\cA_{\pm} \equiv \sum_{j\geq 0} \( a^{(0)}_{\pm j} + a^{(+)}_{\pm j} \cD_{+} +
a^{(-)}_{\pm j} \cD_{-} + a^{(1)}_{\pm j} \cD_{+}\cD_{-}\)\pa^{\pm j}
\lab{N2-oper}
\ee
where the coefficients are $N\!\! =\!\! 2$ superfields, \textsl{i.e.} functions of
$(x,\th_{+},\th_{-})$ and, possibly, of additional (time-evolution) parameters.
The subscripts $(\pm )$ denote the purely differential or purely
pseudo-differential parts of $\cA$, respectively. The rules of conjugation 
within the super-pseudo-differential formalism are:
% as follows (cf. first ref.\ct{SKP-Hen}):
$(\cA \cB )^\ast = (-1)^{|A|\, |B|} \cB^\ast \cA^\ast$
for any two elements with Grassmann parities $|A|$ and $|B|$;
$\(\pa^k\)^\ast = (-1)^k \pa^k\, ,\,\(\cD_{\pm}^k\)^\ast = (-1)^{k(k+1)/2} \cD_{\pm}^k$
and $u^\ast = u$ for any coefficient superfield. Furthermore, in order to 
avoid confusion we shall also employ the following notations: for any 
super-(pseudo-)\-differential operator $\cA$ and a superfield function $f$, 
the symbol $\, \cA (f)\,$ or $\, (\cA f)\,$ will indicate application (action)
of $\cA$ on $f$, whereas the symbol $\cA f$ without brackets will denote simply
operator product of $\cA$ with the zero-order (multiplication) operator $f$.

In ref.\ct{n2-skp} (see also \ct{N2-SKP-other}) a general class \cSKPe of reduced
$N\!\! =\!\! 2$ super-KP integrable hierarchies has been proposed, described by
the following {\em fermionic} $N\!\! =\!\! 2$ super-pseudo-differential Lax 
operators:
\br
% \cL \equiv \cL_{(M_B,M_F)} = 
\cL =\cD_{-} + \sum_{a=1}^{M_B} \P_a \cD_{+}^{-1} \Psi_a +
\sum_{\a=1}^{M_F} \cF_\a \cD_{+}^{-1} \cG_\a
\lab{Lax-SKP-N2}\\
\equiv \cD_{-} + \sum_{i=1}^{M} \P_i \cD_{+}^{-1} \Psi_i \quad ,\quad
M\equiv M_B + M_F
\nonu
\er
Here $\bigl\{\P_a,\Psi_a\bigr\}$ are bosonic, whereas
$\bigl\{\cF_\a,\cG_\a\bigr\}$ are fermionic coefficient superfields.
In what follows we will often use the short-hand notations
$\bigl\{\P_i\bigr\}_{i=1}^{M} \equiv \bigl(\bigl\{\P_a\bigr\}_{a=1}^{M_B},\,
\bigl\{\cF_\a\bigr\}_{\a=1}^{M_F}\bigr)$ and
$\bigl\{\Psi_i\bigr\}_{i=1}^{M} \equiv \bigl(\bigl\{\Psi_a\bigr\}_{a=1}^{M_B},\,
\bigl\{\cG_\a\bigr\}_{\a=1}^{M_F}\bigr)$. Also we will need
the explicit expressions of $\(\cL^K\)_{-}$ for arbitrary odd integer power $K$ 
of $\cL$ \ct{match,N2-SKP-other,n2-skp}:
% \be
% \(\cL^{2k}\)_{-} = \sum_{i=1}^M \sum_{s=0}^{2k-1} (-1)^{s|i|}
% \cL^{2k-1-s}(\P_i) \cD_{+}^{-1} \(\cL^s\)^\ast (\Psi_i)
% \lab{Lax-SKP-N2-2k}
% \ee
\br
\(\cL^{2k+1}\)_{-} = \sum_{i=1}^M \sum_{s=0}^{2k} (-1)^{s|i|}
\cL^{2k-s}(\P_i) \cD_{+}^{-1} \(\cL^s\)^\ast (\Psi_i)
\nonu \\
+ \sum_{i=1}^M \sum_{s=0}^{2k-1} (-1)^{s|i|+s+|i|}
\cL^{2k-1-s}(\P_i) \cD_{+}^{-1}\cD_{-} \(\cL^s\)^\ast (\Psi_i)
\lab{Lax-SKP-N2-2k+1}
\er
and similarly for even powers $K=2k$.
The $N\!\! =\!\! 2$ super-Lax operator \rf{Lax-SKP-N2} can be alternatively
represented as $\cL = \cW \cD_{-}\cW^{-1}$, in terms of the $N\!\! =\!\! 2$ Sato
super-dressing operator $\cW = 1 + \sum_{j\geq 1} w_{j/2} \cD_{+}^{-j}$
whose coefficients superfields $w_{j/2}$ are recursively expressed through
the finite number of the super-Lax coefficient superfields
$\bigl\{\P_i,\Psi_i\bigr\}_{i=1}^M$.

The \cSKPe integrable hierarchies are given by the infinite sets of
bosonic isospectral (w.r.t. $\partder{}{t_l}$-flows with $l\! =\! 1,2,\ldots$)
and fermionic isospectral (w.r.t. $D^{\pm}_n$-flows with $N\!\! =\!\! 1,2,\ldots$) 
Sato evolution equations with $\cL$ as in \rf{Lax-SKP-N2}:
\br
\partder{}{t_l} \cL = \Sbr{\(\cL^{2l}\)_{+}}{\cL}   \quad ,\quad
D^{+}_n \cL = \lcurl \(\L^{2n-1}\)_{+},\,\cL \rcurl
\nonu \\
D^{-}_n \cL = - \lcurl \(\cL^{2n-1}\)_{-} - X_{2n-1},\,\cL \rcurl 
\lab{SKP-N2-isospec-eqs}
\er
where $\L \equiv \cW \cD_{+}\cW^{-1}$ and where (cf. Eq.\rf{Lax-SKP-N2-2k+1}):
\br
\(\cL^{2n-1}\)_{-} - X_{2n-1} \equiv 
\nonu \\
\sum_{i=1}^M \sum_{s=0}^{2n-2}
(-1)^{s(|i|+1)} \cL^{2n-2-s}(\P_i) \cD_{+}^{-1} \(\cL^s\)^\ast (\Psi_i)
\lab{X-def}
\er
The fermionic isospectral flows $D^{\pm}_n$ in Eqs.\rf{SKP-N2-isospec-eqs} 
possess natural realization in terms of two infinite sets of anticommuting 
``evolution'' parameters $\lcurl \rho^{\pm}_n\rcurl_{n=1}^\infty$
and span $N\!\! =\!\! 2$ supersymmetry algebra :
\br
D^{\pm}_n = \partder{}{\rho^{\pm}_n} - 
\sum_{k=1}^\infty \rho^{\pm}_k \partder{}{t_{n+k-1}}
\nonu \\
\lcurl D^{\pm}_n,\, D^{\pm}_m \rcurl = - 2 \partder{}{t_{n+m-1}} \quad ,\quad
\lcurl D^{\pm}_n,\, D^{\mp}_m \rcurl = 0
\lab{N2-isospec-alg}
\er
the rest of flow commutators being zero.

Accordingly, the superfields $\P_i$ and $\Psi_i$ entering the
pseudo-differential art of $\cL$ \rf{Lax-SKP-N2} obey the following infinite
set of bosonic and fermionic nonlinear evolution equations:
\be
\partder{}{t_l}\P_i = \(\cL^{2l}\)_{+}(\P_i) \quad ,\quad
\partder{}{t_l}\Psi_i = -\(\cL^{2l}\)^\ast_{+}(\Psi_i)
\lab{bos-SEF-eqs}
\ee
\be
D^{-}_n \P_i = \Bigl(\cL^{2n-1}_{+} + X_{2n-1}\Bigr) (\P_i) - 2\cL^{2n-1}(\P_i)
\lab{Dn-SEF-eqs}
\ee
\be
D^{-}_n \Psi_i = - \Bigl(\(\cL^{2n-1}\)^\ast_{+} + X^\ast_{2n-1}\Bigr) (\Psi_i)
+ 2\(\cL^{2n-1}\)^\ast (\P_i)
\lab{Dn-adj-SEF-eqs}
\ee
\be
D^{+}_n \P_i = \(\L^{2n-1}\)_{+}(\P_i) \quad ,\quad
D^{+}_n \Psi_i = - \(\L^{2n-1}\)^\ast_{+}(\Psi_i)
\lab{Dn+SEF-eqs}
\ee
Henceforth, all superfield functions pertinent to the integrable \cSKPe
hierarchies depend on $(x,\th_{\pm}; \underline{t},\underline{\rho}^{\pm})$
where the collective notations $\underline{t} \equiv (t_2,t_3,\ldots)$ and
$\underline{\rho}^{\pm} \equiv (\rho^{\pm}_1,\rho^{\pm}_2,\ldots)$ are employed.

All solutions of \cSKPe hierarchies \rf{bos-SEF-eqs}--\rf{Dn+SEF-eqs} are
expressed through a single $N\!\! =\!\! 2$ {\em super-tau function}
~$\t = \t (x,\th_{\pm}; \underline{t},\underline{\rho}^{\pm})$. The latter
is related to the coefficients of the pertinent $N\!\! =\!\! 2$ super-Lax operator 
$\cL = \cW \cD_{-}\cW^{-1}$ \rf{Lax-SKP-N2} and its associate 
$\L = \cW \cD_{+}\cW^{-1}$ as follows \ct{n2-skp} :
\br
\(\cL^{2k}\)_{(-1)} = \partder{}{t_k}\cD_{+} \ln\t  \quad ,\quad
\(\L^{2n-1}\)_{(-1)} = D^{+}_n \cD_{+} \ln\t
\nonu \\
\(\cL^{2n-1} - X_{2n-1}\)_{(-1)} = D^{-}_n \cD_{+} \ln\t 
\phantom{aaaaaaaaa}
\lab{N2-supertau}
\er
where the subscript $(-1)$ indicates taking the coefficient in front of
$\cD_{+}^{-1}$ in the expansion of the corresponding super-pseudo-differential
operator. 
% The proof of the validity of Eqs.\rf{N2-supertau} follows directly 
% from the $N\!\! =\!\! 2$ superspace Zakharov-Shabat ``zero-curvature'' equations, 
% \textsl{i.e.}, the compatibility conditions among the isospectral Sato evolution
% equations \rf{SKP-N2-isospec-eqs} respecting the (anti-)commutation isospectral
% flow algebra \rf{N2-isospec-alg}. 

%
\section{New Ordinary Integrable Hierarchies as Bosonic Limits of Super-KP 
Hierarchies}
\label{sec:2}

Let us first consider the simpler case of reduced super-KP integrable 
hierarchies \cSKP \ct{match} in $N\!\! =\!\! 1$
superspace $(x,\th)$ defined by super-Lax operators:
\br
\cL = \cD + \cF_0 + \sum_{a=1}^{M_B} \P_a \cD^{-1} \Psi_a +
\sum_{\a=1}^{M_F} \cF_a \cD^{-1} \cG_a
\lab{Lax-SKP-N1} \\
\cD \cF_0 = 2\Bigl(\sum_\a \cF_\a \cG_\a - \sum_a \P_a \Psi_a\Bigr)
\er
\br
\partder{}{t_l} \cL = \Sbr{\(\cL^{2l}\)_{+}}{\cL}   \phantom{aaaaaa}
\lab{SKP-N1-isospec-eqs} \\
D_n \cL = - \lcurl \(\cL^{2n-1}\)_{-} - X_{2n-1},\,\cL \rcurl 
\nonu
\er
Here $\cD\!\! =\!\! \pa/\pa\th\! +\! \th \pa/\pa x$ is the single $N\!\! =\!\! 1$
super-covariant derivative; the single set of fermionic isospectral flows $D_n$
are of the same form as their $N\!\! =\!\! 2$ counterparts \rf{N2-isospec-alg} by 
identifying $\rho^{\pm}_n = \rho_n$, and similarly for 
$\(\cL^{2n-1}\)_{-} - X_{2n-1}$ by replacing $\cD_{+}$ with $\cD$ in the 
corresponding $N\!\! =\!\! 2$ counterpart \rf{X-def}. The super-Lax \rf{Lax-SKP-N1} 
coefficient $N\!\! =\!\! 1$ superfields $\bigl\{\P_a,\, \Psi_a\bigr\}$ and 
$\bigl\{\cF_\a,\, \cG_\a\bigr\}$ are bosonic and fermionic, respectively, as in
\rf{Lax-SKP-N2}. The superspace component expansion of the latter reads:
\be
\P_a (x,\th) = u_a (x) + \th f_a (x)  \;\; ,\;\;
\Psi_a (x,\th) = {\bar u}_a (x) + \th {\bar f}_a (x)
\lab{N1-exp-bos}
\ee
\be
\cF_\a (x,\th) = g_\a (x) + \th v_\a (x)  \;\; ,\;\;
\cG_\a (x,\th) = {\bar g}_\a (x) + \th {\bar v}_\a (x)
\lab{N1-exp-fer}
\ee
where $\bigl\{ u_a,{\bar u}_a, v_\a, {\bar v}_\a\bigr\}$ are ordinary bosonic
fields while $\bigl\{ f_a,{\bar f}_a, g_\a, {\bar g}_\a\bigr\}$ are ordinary 
fermionic (anti-commuting) fields. In Eqs.\rf{N1-exp-bos}--\rf{N1-exp-fer} 
we have skipped the dependence on the ``time''-evolution
parameters of the underlying integrable hierarchy \rf{SKP-N1-isospec-eqs}. 

Let us now consider the lowest nontrivial evolution equations of the
$N\!\! =\!\! 1$ super-KP system \rf{Lax-SKP-N1}--\rf{SKP-N1-isospec-eqs}
(cf. Eqs.\rf{bos-SEF-eqs}):
\br
\partder{}{t_2} \P_a = \(\cL^4\)_{+} (\P_a) \quad ,\quad
\partder{}{t_2} \Psi_a = - \(\cL^4\)^\ast_{+} (\Psi_a)
\lab{bos-SEF-eqs-N1} \\
\partder{}{t_2} \cF_\a = \(\cL^4\)_{+} (\cF_\a) \quad ,\quad
\partder{}{t_2} \cG_\a = - \(\cL^4\)^\ast_{+} (\cG_\a)
\lab{fer-SEF-eqs-N1}
\er
and let us insert above the superspace component expansions 
\rf{N1-exp-bos}--\rf{N1-exp-fer}. In the bosonic limit, \textsl{i.e.}, when all
anti-commuting component fieds are set to zero,
Eqs.\rf{bos-SEF-eqs-N1}--\rf{fer-SEF-eqs-N1} reduce to the following system
of nonlinear evolution equations for the bosonic component fields
$\bigl\{ u_a,{\bar u}_a \bigr\}_{a=1}^{M_B}$ and
$\bigl\{ v_\a, {\bar v}_\a \bigr\}_{\a=1}^{M_F}$ :
\br
\partder{}{t_2} u_a = \pa^2 u_a + 2 Q(u,{\bar u},v,{\bar v}) u_a
\nonu \\
\partder{}{t_2} {\bar u}_a = - \pa^2 {\bar u}_a - 
2 {\bar Q} (u,{\bar u},v,{\bar v}) {\bar u}_a
\lab{u-eqs} \\
\partder{}{t_2} v_\a = \pa^2 v_\a + 2 {\bar Q}(u,{\bar u},v,{\bar v}) v_\a
\nonu \\
\partder{}{t_2} {\bar v}_\a = - \pa^2 {\bar v}_\a -
2 Q(u,{\bar u},v,{\bar v}) {\bar v}_\a
\lab{v-eqs}
\er
where:
\br
Q(u,{\bar u},v,{\bar v}) \equiv \sum_{\b=1}^{M_F} v_\b {\bar v}_\b 
- \sum_{b=1}^{M_B} u_b (\pa{\bar u}_b) 
- \Bigl(\sum_{b=1}^{M_B} u_b {\bar u}_b\Bigr)^2 \phantom{aa}
\lab{def-Q} \\
{\bar Q}(u,{\bar u},v,{\bar v}) \equiv \sum_{\b=1}^{M_F} v_\b {\bar v}_\b 
+\sum_{b=1}^{M_B} (\pa u_b) {\bar u}_b 
- \Bigl(\sum_{b=1}^{M_B} u_b {\bar u}_b\Bigr)^2 \phantom{aa}
\lab{def-Q-bar}
\er
From Eqs.\rf{u-eqs}--\rf{def-Q-bar} we conclude that $N\!\! =\!\! 1$ super-KP
hierarchies \rf{Lax-SKP-N1}--\rf{SKP-N1-isospec-eqs} contain in the purely 
bosonic limit new types of ordinary (non-supersymmetric) integrable hiearchies.
The latter are systems of $M_F$-component nonlinear Schr{\"o}\-din\-ger
hierarchies, given by the fields $\bigl\{ v_\a, {\bar v}_\a \bigr\}$,
coupled to $M_B$-com\-po\-nent derivative nonlinear Schr{\"o}\-din\-ger
hierarchies given by the fields $\bigl\{ u_a,{\bar u}_a \bigr\}$ in 
the Ger\-dji\-kov-Ivanov \ct{Gerdji} form.

We can now straightforwardly generalize the above discussion to the $N\!\! =\!\! 2$
super-KP case. The lowest nontrivial evolution equations for the \cSKPe
hierarchy \rf{SKP-N2-isospec-eqs} have the same form as 
\rf{bos-SEF-eqs-N1}--\rf{fer-SEF-eqs-N1} where now $\cL$ is given
by \rf{Lax-SKP-N2}, whereas the $N\!\! =\!\! 2$ superspace component expansions
for the pertinent superfields now read (cf. \rf{N1-exp-bos}--\rf{N1-exp-fer}) :
\br
\P_a (x,\th_{+},\th_{-}) = u_a (x) + \sum_{\pm} \th_{\pm} f^{(\pm)}_a (x)
+ \th_{+}\th_{-} w_a (x) \phantom{aa} 
\nonu \\
\Psi_a (x,\th_{+},\th_{-}) = {\bar u}_a (x) +
\sum_{\pm} \th_{\pm} {\bar f}^{(\pm)}_a (x) + \th_{+}\th_{-} {\bar w}_a (x)
\phantom{aa} \lab{N2-exp-bos} \\
\cF_\a (x,\th_{+},\th_{-}) = f_\a (x) + \sum_{\pm} \th_{\pm} v^{(\pm)}_\a (x)
+ \th_{+}\th_{-} g_\a (x)  \phantom{aa} 
\nonu \\
\cG_\a (x,\th_{+},\th_{-}) = {\bar f}_\a (x) + 
\sum_{\pm} \th_{\pm} {\bar v}^{(\pm)}_\a (x) + \th_{+}\th_{-} {\bar g}_\a (x)
\phantom{aa} \lab{N2-exp-fer}
\er
where we have suppressed the ``time''-evolution dependence for brevity. Here
$\bigl\{\st{-}{u_a}, \st{-}{w_a}, \st{-}{v}\!\!{}^{(\pm)}_\a \bigr\}$ are
ordinary bo\-so\-nic fields whereas
$\bigl\{\st{-}{f_\a},\st{-}{g_\a}, \st{-}{f}\!\!{}^{(\pm)}_a\bigr\}$
are ordinary fermionic (anti-commuting) fields. Inserting the superspace
expansions \rf{N2-exp-bos}--\rf{N2-exp-fer} in
Eqs.\rf{bos-SEF-eqs-N1}--\rf{fer-SEF-eqs-N1} with $\cL$ given by
\rf{Lax-SKP-N2} we obtain:
\br
\partder{}{t_2} u_a = \pa^2 u_a + 
2 Q\bigl(\st{-}{u},\st{-}{w},\st{-}{v}\bigr) u_a
\nonu \\
\partder{}{t_2} {\bar u}_a = - \pa^2 {\bar u}_a - 
2 {\bar Q}\bigl(\st{-}{u},\st{-}{w},\st{-}{v}\bigr) {\bar u}_a
\lab{u-eqs-2} \\
\partder{}{t_2} w_\a = \pa^2 w_\a + 
2 {\bar Q}\bigl(\st{-}{u},\st{-}{w},\st{-}{v}\bigr) w_\a
\nonu \\
- 2 \Bigl(\pa\bigl(\sum_b u_b {\bar u}_b\bigr)\Bigr) \pa u_a
\nonu \\
\partder{}{t_2} {\bar w}_\a = - \pa^2 {\bar w}_\a -
2 Q\bigl(\st{-}{u},\st{-}{w},\st{-}{v}\bigr) {\bar w}_\a
\nonu \\
- 2 \Bigl(\pa\bigl(\sum_b u_b {\bar u}_b\bigr)\Bigr) \pa {\bar u}_a
\lab{w-eqs-2} \\
\partder{}{t_2} v^{(+)}_\a = \pa^2 v^{(+)}_\a  + 
2 {\bar Q}\bigl(\st{-}{u},\st{-}{w},\st{-}{v}\bigr) v^{(+)}_\a 
\nonu \\
\partder{}{t_2} {\bar v}^{(+)}_\a = - \pa^2 {\bar v}^{(+)}_\a  - 
2 Q\bigl(\st{-}{u},\st{-}{w},\st{-}{v}\bigr) {\bar v}^{(+)}_\a 
\lab{v+eqs-2} \\
\partder{}{t_2} v^{(-)}_\a = \pa^2 v^{(-)}_\a  + 
2 Q\bigl(\st{-}{u},\st{-}{w},\st{-}{v}\bigr) v^{(-)}_\a 
\nonu \\
+ 2 \Bigl(\pa\bigl(\sum_b u_b {\bar u}_b\bigr)\Bigr) v^{(+)}_\a 
\nonu \\
\partder{}{t_2} {\bar v}^{(-)}_\a = - \pa^2 {\bar v}^{(-)}_\a  - 
2 {\bar Q}\bigl(\st{-}{u},\st{-}{w},\st{-}{v}\bigr) {\bar v}^{(-)}_\a 
\nonu \\
+ 2 \Bigl(\pa\bigl(\sum_b u_b {\bar u}_b\bigr)\Bigr) {\bar v}^{(+)}_\a 
\lab{v-eqs-2}
\er
where:
\be
Q\bigl(\st{-}{u},\st{-}{w},\st{-}{v}\bigr) \equiv 
\sum_{\b=1}^{M_F} v^{(-)}_\b {\bar v}^{(+)}_\b
+ \sum_{b=1}^{M_B} u_b {\bar w}_b 
+ \Bigl(\sum_{b=1}^{M_B} u_b {\bar u}_b\Bigr)^2
\lab{def-Q-2}
\ee
\be
{\bar Q}\bigl(\st{-}{u},\st{-}{w},\st{-}{v}\bigr) \equiv 
\sum_{\b=1}^{M_F} v^{(+)}_\b {\bar v}^{(-)}_\b
- \sum_{b=1}^{M_B} w_b {\bar u}_b
+ \Bigl(\sum_{b=1}^{M_B} u_b {\bar u}_b\Bigr)^2 
\lab{def-Q-bar-2}
\ee
Eqs.\rf{u-eqs-2}--\rf{def-Q-bar-2} bring us to the conclusion that $N\!\! =\!\! 2$
super-KP hierarchies \rf{Lax-SKP-N2},\rf{SKP-N2-isospec-eqs} contain in the
purely bosonic limit new types of ordinary non-supersymmetric integrable
hierarchies. The latter are coupled systems of several {\em multi-component} 
non-linear Schr{\"o}dinger-type hierarchies whose basic nonlinear evolution
equations \rf{u-eqs-2}--\rf{def-Q-bar-2} contain {\em additional} (besides
the usual cubic terms) quintic and {\em higher-derivative} nonlinear terms.

Finally, let us note that $N\!\! =\!\! 1,2$ super-KP hierarchies possess a vast
set of {\em additional non-isospectral} symmetries which span
infinite-dimensional non-Abelian superloop superalgebras \ct{svirflow,n2-skp}.
\section{$N\!\! =\!\! 2$ Super-Toda Chain}
\label{sec:3}
\DB (DB) transformations for $N\!\!\! =\!\!\! 1,2$ super-KP hierarchies have been worked
% \DB (DB) transformations for super-KP hierarchies have been worked
out in detail in refs.\ct{match,zim-ber,svirflow,n2-skp}, where we have
derived the explicit form of the general DB (``super-soliton''-like) solutions.
The latter are given in terms of Berezinians (super-deteminants) whose
bosonic and fermionic blocks have a special generalized Wronskian-like
structure. 

Here we will discuss in some detail the DB orbit, \textsl{i.e.}, the
sequence of successive iterations of DB transformations for the simplest
$N\!\! =\!\! 2$ super-KP hierarchy $\textsl{SKP}^{N=2}_{(1,0)}$
with super-Lax operator $\cL = \cD_{-} + \P \cD_{+}^{-1} \Psi$ :
% \br
% \cL = \cD_{-} + \P \cD_{+}^{-1} \Psi  \phantom{aaaaaaaaaaaa}
% \lab{Lax-SKP-N2-1} \\
\br
\Psi^{(n+1)} = {1\o {\P^{(n)}}} \quad ,\quad
\P^{(n+1)} = - \P^{(n)} \cD_{+}\Bigl(\frac{\cL^{(n)}(\P^{(n)})}{\P^{(n)}}\Bigr)
\nonu \\
= - \P^{(n)}\cD_{+}\cD_{-}\ln \P^{(n)} - \bigl(\P^{(n)}\bigr)^2 \Psi^{(n)}
\phantom{aa}
\lab{DB-orbit}
\er
where the subscripts in brackets indicate the number of iteration steps of DB
transformations. Formulas \rf{DB-orbit} are simple special case of the general
expressions for successive DB transformations \ct{match,zim-ber,svirflow,n2-skp}.
Introducing new $N\!\! =\!\! 2$ superfields $\vp_n$ through the substitution 
$\P^{(n)} = e^{\vp_n}$, we can rewrite Eqs.\rf{DB-orbit} in the following form:
\be
\cD_{-}\cD_{+} \vp_n = e^{\vp_{n+1}-\vp_{n}} + e^{\vp_{n}-\vp_{n-1}}
\lab{N2-Super-Toda}
\ee
which is the $N\!\! =\!\! 2$ supersymmetric extension of the equations of the
ordinary Toda chain model (for the $N\!\! =\!\! 1$ super-Toda chain see
\ct{match}; for alternative representations of $N\!\! =\!\! 2$ super-Toda chain
see \ct{Sorin-Lecht}). Indeed, inserting in \rf{N2-Super-Toda} the superspace 
component expansion
$\vp_n (x,\th_{+},\th_{-}) = u_n (x) + \sum_{\pm} \th_{\pm} f^{(\pm)}_n +
\th_{+}\th_{-} w_n (x)$, we obtain in the bosonic limit ($f^{(\pm)}_n = 0$)
the following equations for $u_n (x)$ :
\be
-\pa^2 u_n = e^{u_{n+2}-u_{n}} - e^{u_{n}-u_{n-2}}
\lab{Toda-double}
\ee
which are precisely the ordinary Toda-chain equations for {\em double}
Toda lattice spacing.

Using the general Berezinian expressions for the super-tau functions of
$N\!\! =\!\! 2$ super-KP hierarchies \ct{n2-skp} we obtain the following
explicit solution for the $N\!\! =\!\! 2$ super-Toda chain model
\rf{N2-Super-Toda} : $\vp_n = \ln \t^{(n+1)} + \ln \t^{(n)}$
% \be
% \vp_n = \ln \t^{(n+1)} + \ln \t^{(n)}
% \lab{N2-Super-Toda-sol}
% \ee
where the super-tau functions are given by
\br
\(\t^{(2m)}\)^{-1} = 
\mathrm{Ber} \twomat{W_{m,m}(\P_0)}{W_{m,m}(\cD_{-}\P_0)}{
W_{m,m}(\cD_{+}\P_0)}{W_{m,m}(\cD_{+}\cD_{-}\P_0)}  \phantom{aa}
\lab{tau-2m} \\
\t^{(2m+1)} = 
\mathrm{Ber} \twomat{W_{m+1,m+1}(\P_0)}{W_{m,m+1}(\cD_{-}\P_0)}{
W_{m+1,m}(\cD_{+}\P_0)}{W_{m,m}(\cD_{+}\cD_{-}\P_0)} \phantom{aa}
\lab{tau-2m+1}
\er
In Eqs.\rf{tau-2m}--\rf{tau-2m+1} the following notations are used.
$W_{k,m} (F)$ is $k\times m$ rectangular matrix block of the form~
$W_{k,m}(F) =\left\Vert\pa^{i+j-2} F\right\Vert_{i=1,\ldots,m}^{j=1,\ldots,k}$
for any superfield $F(x,\th_{\pm})$. $\P_0$ is explicitly given by
$N\!\! =\!\! 2$ superspace Fourier integral:
\be
\P_0 (x,\th_{\pm}) = \int d\l\,d\eta_{\pm}\, \p_0 (\l,\eta_{\pm})\,
\exp\bigl\{\l x + \sum_{\pm} \eta_{\pm} \th_{\pm} \bigr\}
\lab{P-0}
\ee
with an arbitrary $N\!\! =\!\! 2$ superspace ``density''
$\phi_0 (\l,\eta_{\pm}) = \phi^{(1)}_B (\l) + \eta_{+} \phi^{(1)}_F (\l)
+ \eta_{-} \bigl( \phi^{(2)}_F (\l) + \eta_{+} \phi^{(2)}_B (\l)\bigr)$
where $\eta_{\pm}$ are anti-commuting ``Fourier momenta''.
%
% \section{Outlook}
% \label{sec:4}

It is a very interesting topic for further research to study in more details
the properties and possible physical significance of the very broad class of
DB (``super-soliton''-like) solutions of $N\!\! =\!\! 1,2$ super-KP
integrable hierarchies. 

${}$ \\
\noindent
\textbf{Acknowledgements.}
We gratefully acknowledge the hospitality of the Organizing Committee of
the International Conference {\em ``Geometry, Integrability and
Nonlinearity in Condensed Matter and Soft Condensed Matter Physics''}
(July 2001, Bansko/Bulgaria) where the present results were first reported.
We thank Prof. V. Gerdjikov for illuminating discussions. This work is also 
partially supported by Bulgarian NSF grant \textsl{F-904/99}.

\end{document}